\documentclass[openacc]{rsproca_new}
\usepackage{fix-cm}
\usepackage{lipsum}
\usepackage{multicol}
\usepackage{amsmath,amsfonts,amssymb}
\usepackage{graphicx}
\usepackage{multirow}
\usepackage{dsfont}
\usepackage{xr}

\newcommand*{\addFileDependency}[1]{
 \typeout{(#1)}
 \@addtofilelist{#1}
 \IfFileExists{#1}{}{\typeout{No file #1.}}
}
\makeatother

\jname{rsif}
\Journal{J R Soc Interface\ }

\begin{document}

\title{Scarce Data, Noisy Inferences, and Overfitting: The Hidden Flaws in Ecological Dynamics Modelling}

\author{Mario Castro$^{1,2}$, Rafael Vida$^{1,2,4}$, Javier Galeano$^{2,4}$, Jos\'e A. Cuesta$^{2,3,5}$}

\address{$^{1}$Institute for Research in Technology (IIT), Universidad Pontificia Comillas, Madrid, Spain\\$^{2}$Grupo Interdisciplinar de Sistemas Complejos (GISC), Madrid, Spain\\$^{3}$Universidad Carlos III de Madrid, Departamento de Matem\'aticas, Legan\'es, Spain\\$^{4}$Complex System Group, Universidad Polit\'ecnica de Madrid, Madrid, Spain\\$^{5}$Instituto de Biocomputaci\'on y F\'{\i}sica de Sistemas Complejos, Universidad de Zaragoza, Zaragoza, Spain}

\subject{xxxxx, xxxxx, xxxx}


\corres{Corresponding author\\
\email{marioc@comillas.edu}}





\begin{abstract}

Metagenomic data has significantly advanced microbiome research by employing ecological models, particularly in personalised medicine. The generalised Lotka-Volterra (gLV) model is commonly used to understand microbial interactions and predict ecosystem dynamics. However, gLV models often fail to capture complex interactions, especially when data is limited or noisy. This study critically assesses the effectiveness of gLV and similar models using Bayesian inference and a model reduction method based on information theory. We found that ecological data often leads to non-interpretability and overfitting due to limited information, noisy data, and parameter sloppiness. Our results highlight the need for simpler models that align with the available data and propose a distribution-based approach to better capture ecosystem diversity, stability, and competition. These findings challenge current bottom-up ecological modelling practices and aim to shift the focus toward a Statistical Mechanics view of ecology based on distributions of parameters.
\end{abstract}

\begin{fmtext}

Recent decades have witnessed the central role of the human gut microbiome in health and disease, catalysed by the explosion in metagenomic data \cite{Handelsman:1998,Steele:2005} and led by initiatives such as the Human Gut Project~\cite{human2012framework}. These data have triggered activity to quantify and even predict the role of gut composition in driving so-called personalised medicine: a custom-made intervention based on the patient's specific gut ecosystem and the connection between dysbiosis and disease. 

Mirroring the years succeeding the Human Genome Project~\cite{venter2001interna}, it has been suggested that metagenomic data is not a clear-cut \emph{book of life,} and data should be supplemented by bottom-up approximations~\cite{ball2023life}. Modern metagenomic methods sequence all RNA in a sample, allowing for both taxonomic identification (down to species or strain level) and functional analysis (genes, pathways, metabolic potential). In principle, the collected data allows us to estimate the abundances of different species and, through correlations, provide insights into microbial relationships and the interaction between a certain environment and its role in health and disease.
\end{fmtext}

\maketitle

\begin{multicols}{2}
In parallel with data-based analysis, researchers have increasingly turned to mathematical and computational frameworks rooted in ecology and dynamical systems theory to address this limitation. These frameworks aim to bridge the gap between descriptive data and mechanistic understanding by providing a way to model the interactions within microbial communities. Such models allow for exploring ecosystem stability and resilience and offer predictive power for how interventions or environmental changes may reshape microbial dynamics. These approaches seek to complement metagenomic insights with explanatory and predictive tools by focusing on the principles governing microbial coexistence and interaction.

\begin{figure*}[!htp]
    \centering
    \includegraphics[width=0.85\linewidth]{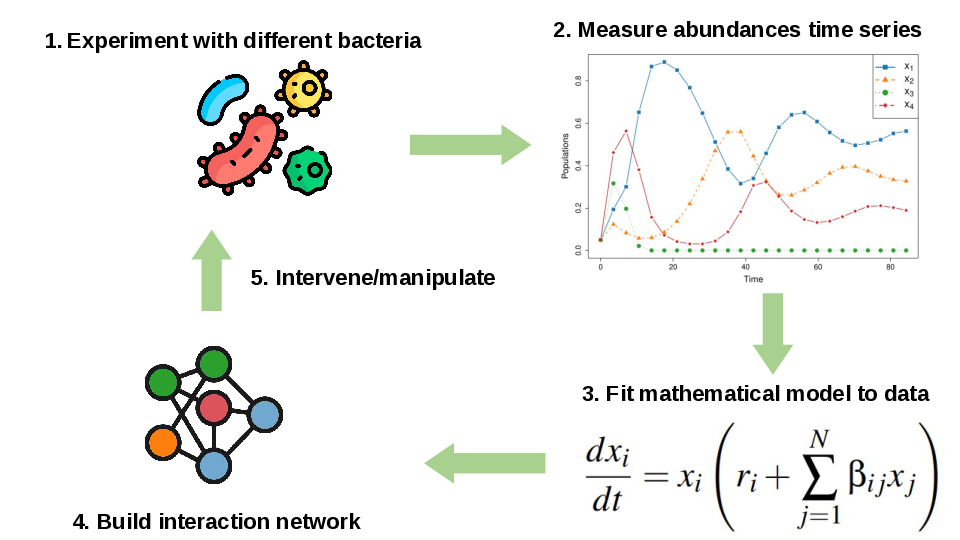}
    \caption{
    \label{fig:cartoon}Sketch illustrating how experiments and models can work synergistically to personalise an individual microbiome.}
\end{figure*}

In particular, the generalised Lotka-Volterra (gLV) model~\cite{venturelli2018deciphering} and its variants (see eq.~\eqref{eq:lotka} for its most basic version) borrowed from theoretical ecology may offer a link between experimental data on bacterial coexistence and microscopic interactions. Hence, the research program proposed by several authors is straightforward: gather data, fit the equations, infer the interaction matrix, and exploit pair-interaction networks~\cite{faust:2012, Stein2013, bucci:2014, jones:2019, matchadoNetworkAnalysisMethods2021, Coyte2021} to predict the outcome of custom-made therapies aimed to reestablish a healthy microbiome~\cite{abreu2018pairing}.
This process is sketched in Fig.~\ref{fig:cartoon}.

In the last few years, there has been evidence about the virtues of the gLV model (mainly in its stochastic variant) to capture universal properties of the microbiome~\cite{faust:2012, bucci:2014, Coyte:2015, jones:2019, grilli2020macroecological, descheemaeker2021, Hu:2022, camacho2024sparse, camacho2024non, mallmin:2024}. However, despite their widespread application, recent literature has begun to cast shadows on the efficacy of these models. Critics argue that the limitations of these models are not merely a matter of practicality concerning model fitting or data availability. Instead, they point to a more profound issue, asserting that these models are inherently flawed in their ability to capture the intricate dynamics of the microbiome~\cite{balsa-cantoMixedGrowthCurve2020, jiMicrobialCommunityDynamics2021, momeniLotkaVolterraPairwiseModeling2017, ramReplyBalsaCantoGrowth2020, pinto2022species, zapien-camposInferringInteractionsMicrobiome2023, picotMicrobialInteractionsTheory2023, LubianaBotelho2025}. Notwithstanding, some of these concerns regarding technical limitations, particularly those related to model fitting, may be mitigated by fitting relative populations, as suggested by \cite{remien2021a}. Other authors argue that the problem lies in the extreme simplicity of Lotka-Volterra equations,  pointing to more sophisticated versions including higher order interactions~\cite{bairey2016high}, or even to include the dynamics of resources that are at the centre of the interaction among species.

In this paper, we use arguments from Bayesian inference and information theory to argue against the viability of this whole bottom-up enterprise. In particular, we show a mismatch between the type of data and the amount of actual information entailed in those data, on the one hand, and the level of description implicit in the gLV model and its variants to capture that information, on the other hand.

Our line of argumentation has three ingredients---that we further develop in the following sections---namely: (i) ecological data capture only a few timescales and individual steady states; (ii) the very structure of the models combined with data acquisition noise provides non-consistent interpretations of the interaction parameters; and finally (iii) gLV-type models supplemented with this sort of data are sloppy---in the sense introduced by~\cite{transtrum2015perspective}---so the parameters overfit the amount of existing information. 

Our main message is that, as traditionally recognised in Statistical Mechanics and Complexity Science, excessive mathematical detail can provide inconsistent explanations of microscopic ecological problems when presented too confidently. We illustrate this point by combining synthetic data, where we have complete knowledge, and analysing the effect of adding noise, which would emulate how experimental data is typically obtained. Overall, we conclude that bottom-up descriptions, such as the generalised Lotka-Volterra (gLV) model and its variants, are often overparameterized. As a result, the interpretation of the fitted coefficients does not align with the actual information contained in the data, claiming a radically different approach to the dynamics and function of the microbiome. Our methodological approach relies on lengthy numerical computations that cannot be automated for many species, so our argumentation is based on counterexamples rather than on mathematically rigorous proofs.

We wrap up this paper arguing that the way out of this flawed enterprise is to adopt the viewpoint of Statistical Mechanics, where the central challenge is to discover appropriate macroscopic state variables that capture the essence of ecological dynamics and formulate parameter distributions compatible with those state variables, akin to the ensembles of Statistical Mechanics. The relevant macroscopic dynamics of microbiomes should follow from those ensembles in the same way that thermodynamics follows from Boltzmann's distribution. In this novel approach, finding the particulars of microbial interactions is no more interesting than determining the specific positions and momenta of particles in a gas.

\section{The three limitations of population dynamics models} %

We will focus our discussion on the gLV model of theoretical ecology, which describes the evolution of the abundances $x_i$ of an ecological community of $i=1,\dots,N$ species through the dynamical system
\begin{equation}
\frac{dx_i}{dt}=x_i\left(r_i+\sum_{j=1}^N\beta_{ij}x_j\right),
\quad i=1\ldots N,
\label{eq:lotka}
\end{equation}
where $r_i$ represents the (unconstrained) exponential growth rate of species $i$ in isolation, and the term $\beta_{ij}x_j$ quantifies the effect of species $j$ in the growth of species $i$. We collectively refer to the coefficients $(\beta_{ij})$ as the interaction matrix. Its diagonal elements can be regarded as (minus) the reciprocal of the so-called carrying capacities of the species in isolation.

We are aware of variants of this model including additional equations for the resources and their consumption (consumer-resource models \cite{MacArthur:1970}), stochastic terms accounting for demographic or environmental fluctuations \cite{camacho2024sparse, camacho2024non}, or terms with higher order interactions \cite{bairey2016high} that aim at capturing the combined effect of several species in the growth rate of another. But, as we will argue in the Discussion, far from fixing the problem we are about to analyse, these extensions only worsen because they proliferate the number of parameters.

\subsection{Data of species abundances contains only a few pieces of information}
\label{ssec:pieces}

As mentioned in the Introduction, we argue that the problem with Eq.~\eqref{eq:lotka} and its extensions to accommodate higher-order interactions or even competition for resources relies not just on the model itself but on the type of data used to infer the parameters of those models.

\begin{figure*}[!htp]
    \centering
    \begin{tabular}{ll}
         \textbf{(A)}&\textbf{(B)}  \\
    \includegraphics[width=.45\textwidth]{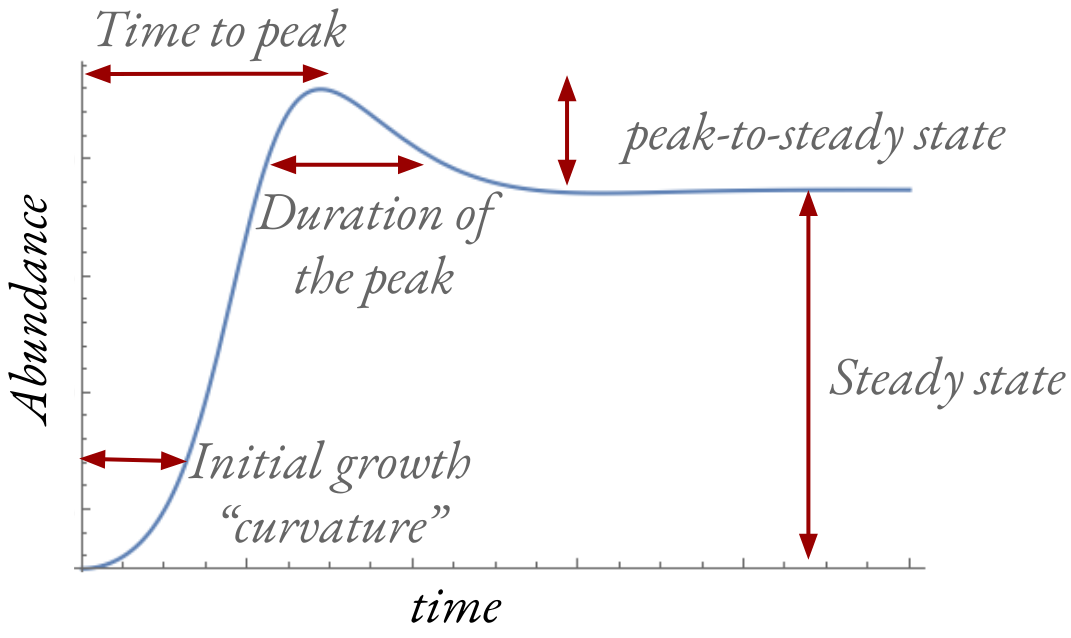}     & 
    \includegraphics[width=.45\textwidth]{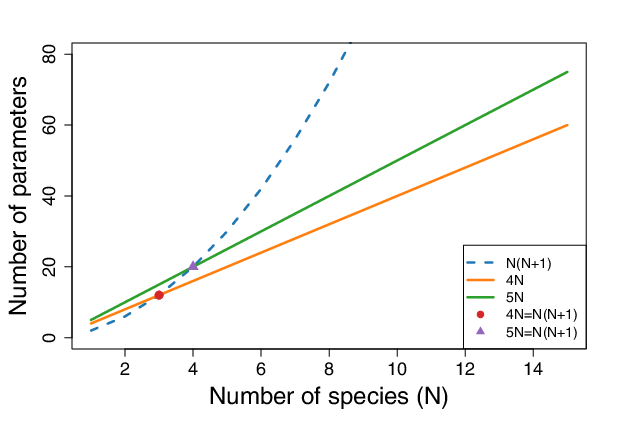}\\
    (C) & (D)\\
          \includegraphics[width=.45\textwidth]{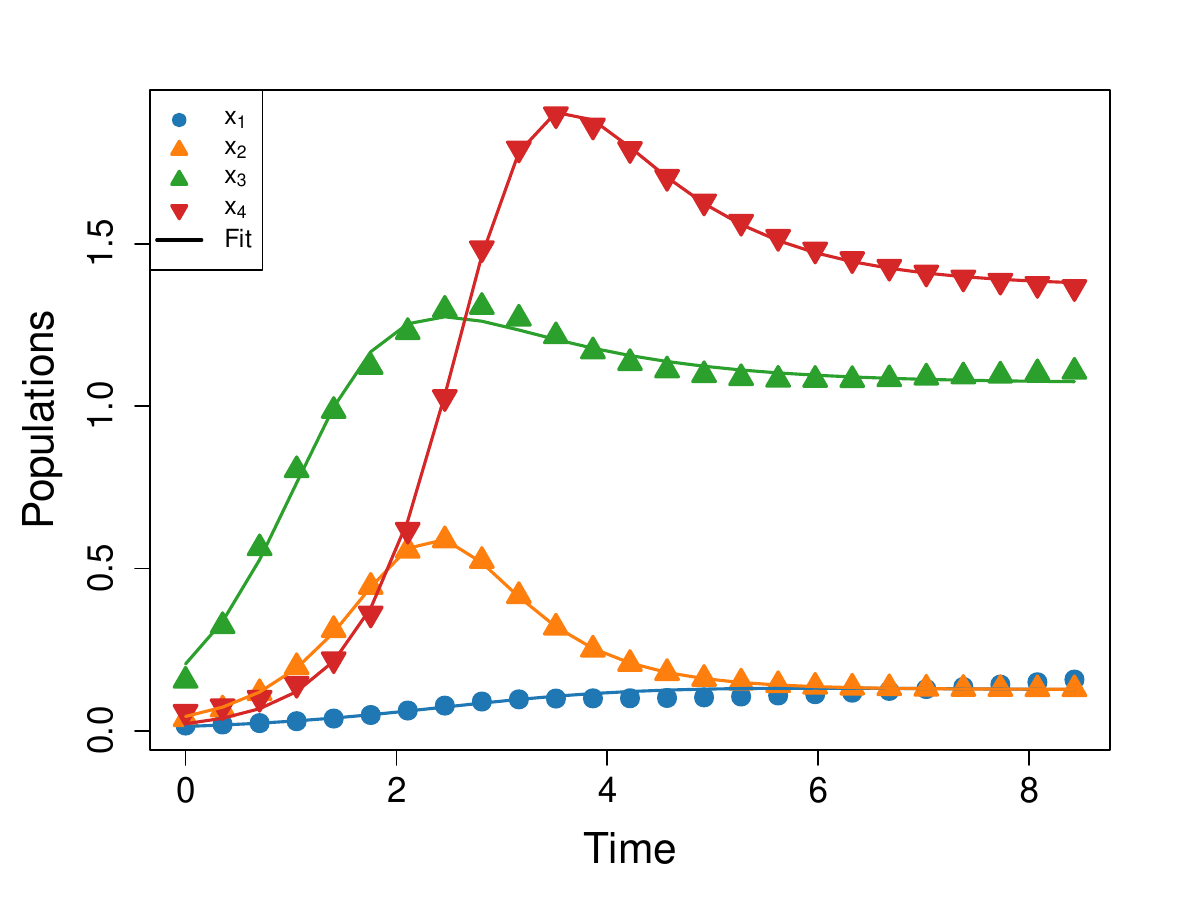} &
          \includegraphics[width=.45\textwidth]{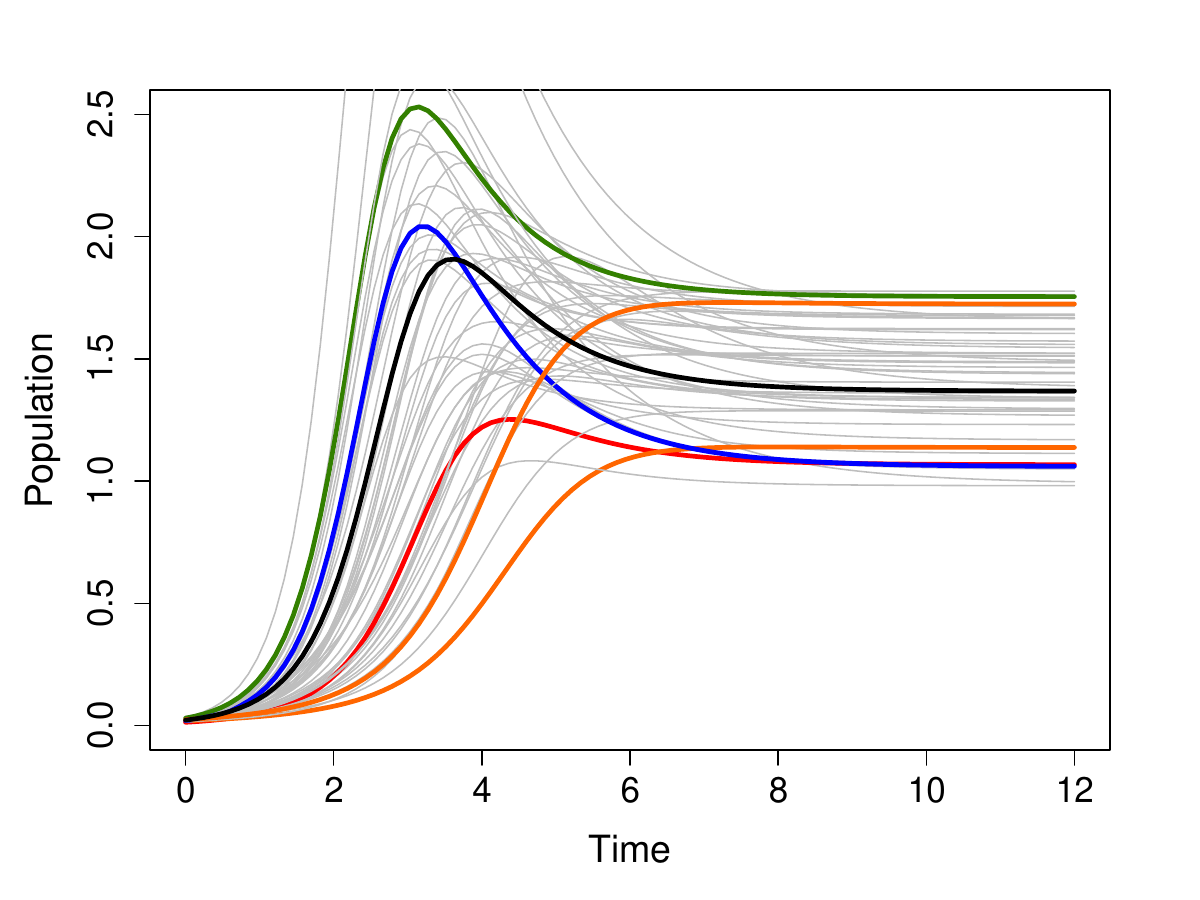} 
          
    \end{tabular}
    \caption{\label{fig:curve}\textbf{(A)} A schematic illustration of how to estimate relevant pieces of information in a time series. As shown, a few parameters can explain this stereotypical shape even if the data has infinite precision (a continuous curve). \textbf{(B)} Number of parameters for the N-species gLV model (dashed blue line) and estimated number of parameters, $4N$--$5N$, (according to the rule of thumb in panel \textbf{(A)}). The symbols show the size above which the number of parameters, while potentially identifiable, is not explainable or meaningful from an ecological perspective.
    \textbf{(C)} An example of the time evolution of the abundance of 4 species according to the gLV equations (symbols). The solid lines on top of each species are fitted Eq.~\eqref{eq:generic}. \textbf{(D)} Illustration of how Eq.~\eqref{eq:generic} accommodates different shapes with just 5 parameters. Colours illustrate how the steady state, peak, and other features can be varied independently. For instance, orange curves (sigmoids) can also be fitted with just 3 parameters.}
\end{figure*}

To illustrate this with a back-of-the-envelope calculation, consider a typical time series, as depicted in Fig.~\ref{fig:curve}(C). The shape of the curve, dissected in Fig.~\ref{fig:curve}(A), suggests that there are at most 4 (5 if one leaves free the initial condition) pieces of independent information per population. For $N$ species, this accounts for $5N$ independent pieces of information.

We can make this claim more quantitative by proposing the (heuristic) equation
\begin{equation}
x(t)=\frac{a_1+a_2e^{-a_3t}}{1+a_4e^{-a_5t}}
    \label{eq:generic}
\end{equation}
to fit curves like that of Fig.~\ref{fig:curve}(A). Figure~\ref{fig:curve}(D) shows how versatile this expression is to reproduce that kind of curve and, in particular, how it describes reasonably well the data of Figure~\ref{fig:curve}(C) (solid lines). Note that the curves can have a sigmoid form, a clearly defined maximum, and different parameters lead to different timescales and peak and steady state locations. Of course, we can fit the same data using formulas with more parameters, but this will overfit the data. Different combinations of the parameters will produce similar fits. This is the fingerprint of sloppiness---parameters can be widely varied without changing the fit appreciably.

In contrast, a model like the gLV equations for $N$ species involves---in its simplest form \eqref{eq:lotka} with at most pairwise interactions---at least $N+N^2$ parameters. The consequence is that for $N\geq 4$, the number of parameters in the model surpasses the actual pieces of information available from the data, as shown in Fig.~\ref{fig:curve}(B), rendering the models inherently sloppy and unreliable for the purpose they are employed, as we discuss below. 

To simulate synthetic ecosystems, we randomly sampled the growth rates $r_i$, as well as the negative self-interactions $-\beta_{ii}$ ($K_i\equiv -\beta_{ii}^{-1}>0$ are carrying capacities) from exponential distributions (the maximum entropy distribution for positive random variables~\cite{jaynes2003probability}) with mean 1/2. Interspecies interactions $\beta_{ij}$ ($i\ne j$) are sampled from normal distributions of mean 0 and standard deviation 1 (the maximum entropy distribution for generic variables with fixed variance). This choice for the priors does not alter the conclusions as they set a typical timescale and the order of magnitude of the abundances at steady-state. After these random numbers are generated, we compute the steady state and stability and iterate until it is feasible (has positive abundances for all species) and linearly stable \cite{may1972will, allesina2012, liuFeasibilityStabilityLarge2023}.

In Fig.~\ref{fig:curve}(C), we show a simulation of 4 species. The dots are the simulation of the gLV equation, and the solid line is the best fit to Eq.~\eqref{eq:generic}. Of course, we are not claiming here that Eq.~\eqref{eq:generic} models the trajectories of the gLV. In fact, if the eigenvalues associated with the stability of the steady state have a non-zero imaginary part, this function would not work---it cannot reproduce oscillatory behaviours. But, we want to rationalise the idea that, disregarding the number of points measured in a curve, the shape of the curve encodes a limited number of parameters (no more than 5 in this case) \emph{regardless of the time-series resolution}.

Note that our argument here is not that the parameters in Eq.~\eqref{eq:lotka} are not practically identifiable~\cite{castro2020testing}; with enough data and precision, they are. We argue that they are not meaningful because they are over-parametrising the data, i.e., there are many more parameters $r_i, \beta_{ij}$ in the model than overall relevant pieces of information in the series. As depicted in Fig.~\ref{fig:curve}(B), we expect this problem to be more dramatic for larger ecosystems because the number of irrelevant parameters scales as $O(N^2)$. 

\subsection{Noisy data provide compatible but contradictory species interactions}
\label{ssec:bayes}

\begin{figure*}[!htp]
    \centering
    \begin{tabular}{ll}
    \includegraphics[width=.5\textwidth]{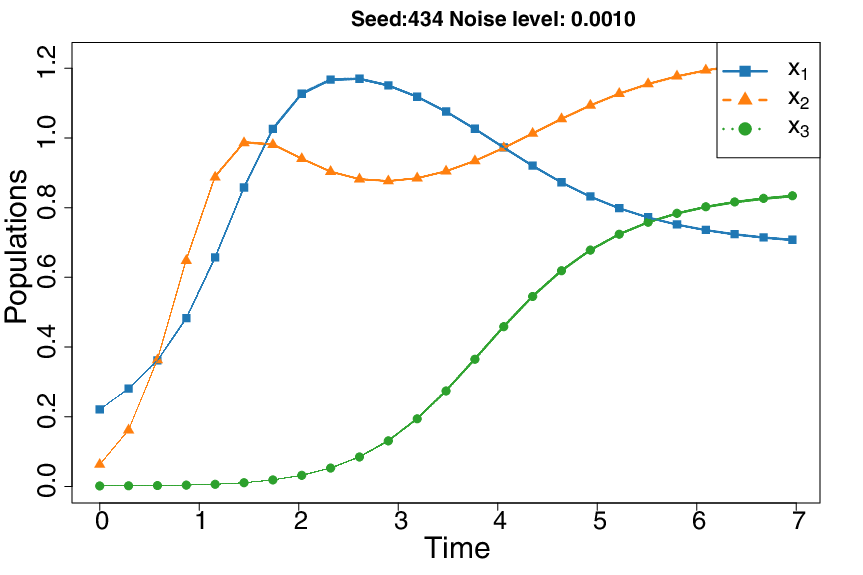} &
    \includegraphics[width=.5\textwidth]{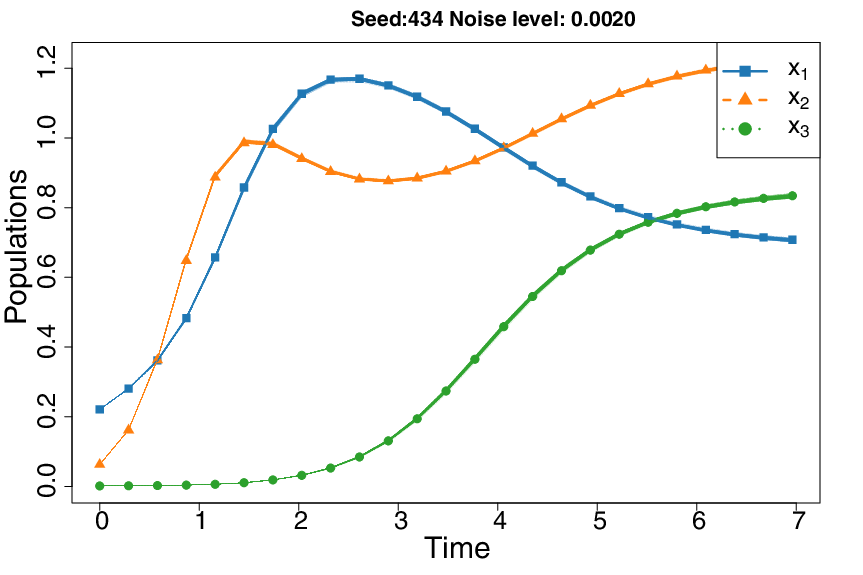}\\
    \includegraphics[width=.5\textwidth]{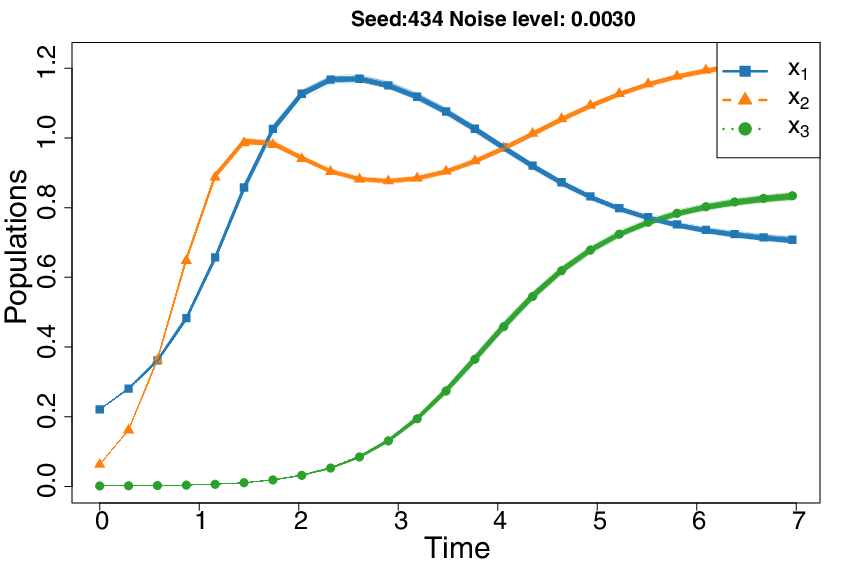}&
    \includegraphics[width=.5\textwidth]{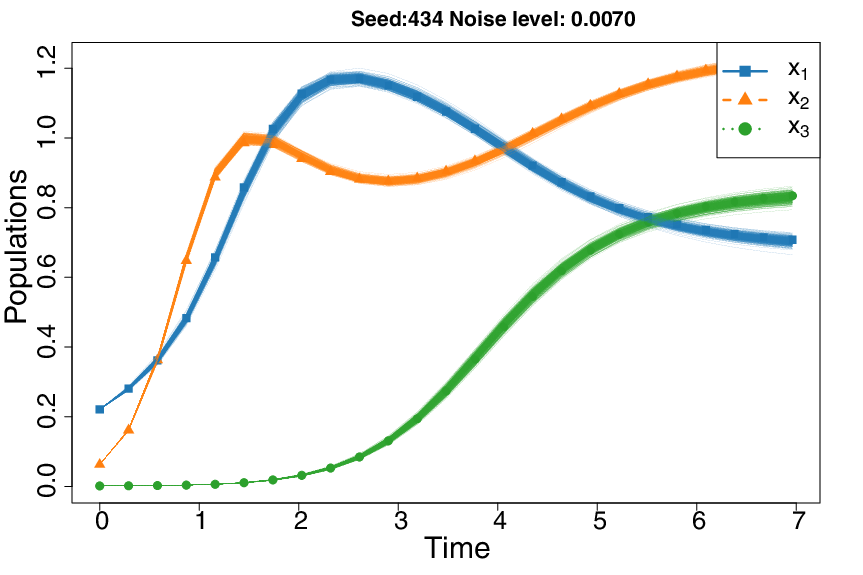}\\
    \includegraphics[width=.5\textwidth]{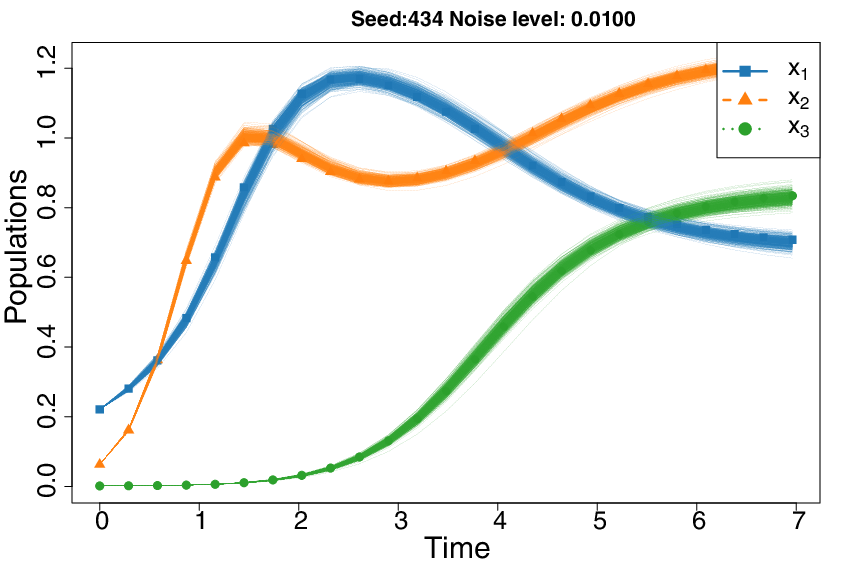}&
    \includegraphics[width=.5\textwidth]{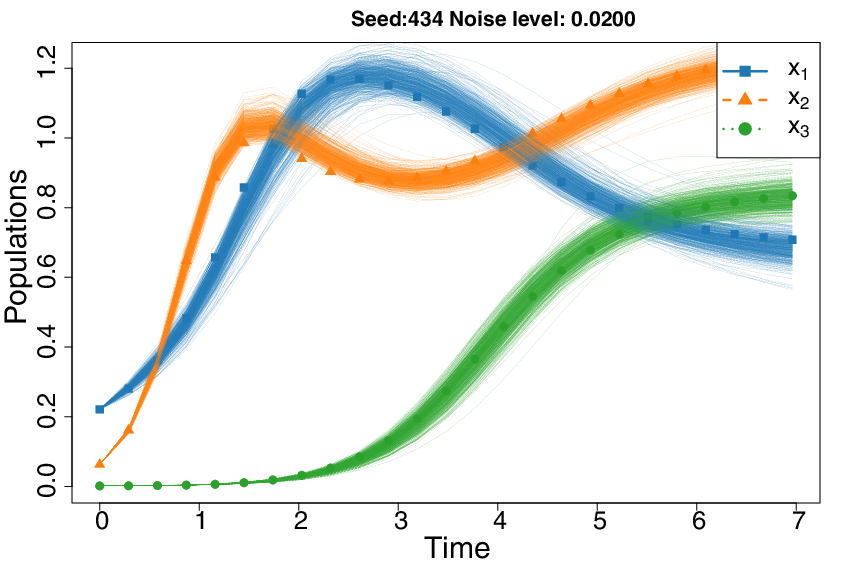}
    \end{tabular}
    \caption{\label{fig:ppc} Posterior predictive trajectories simulated by sampling the posterior distributions (summarized in Sec.~S3) for 6 different log-normal noise levels (title of each panel) added to the deterministic curves in Fig.~\ref{fig:curve}(C). The posterior predictive trajectories capture higher (log-normal) noise levels by increasing the variability for larger abundances.}
\end{figure*}

In this section, we explore the role of noise in parameter inference from data. As Bayesian inference is computationally costly, we illustrate the effect for 3 species (see Fig.~\ref{fig:ppc}) and relegate analogous analyses for 4 and 5 species to the Supplementary Material. 
We have tested this for many random choice parameters, and the conclusions remain the same. . The time series we analyse are obtained by simulating Eq.~\eqref{eq:lotka} and then adding a log-normal noise with different intensities, as this distribution guarantees positivity of the abundances. It is consistent with more complex stochastic versions of the gLV model~\cite{descheemaeker2020stochastic, camacho2024sparse}. We use different values of the log-normal standard deviation to understand the role of noise on inference.

We perform Bayesian inference through the Stan software (see Secs.~S1 and S2 for the details on simulation parameters and the code, also downloadable from Zenodo\footnote{The code is available at \href{https://zenodo.org/records/16747311}{https://zenodo.org/records/16747311}}). This approach allows us to compute the posterior distribution for the \emph{observed} noisy data as
\begin{equation}
P\big(\boldsymbol{\Omega}|\mathbf{y}(t),\mathbf{x}(0)\big)=
\frac{1}{Z}P\big(\mathbf{y}(t)|\mathbf{x}(0),\boldsymbol{\Omega}\big)
P(\boldsymbol{\Omega}),
\label{eq:posterior}
\end{equation}
where boldface letters denote vectors of the corresponding quantities, $\boldsymbol{\Omega}\equiv\{\mathbf{r},\boldsymbol{\beta}, \boldsymbol{\sigma}\}$ (statistically independent),
\begin{equation}
Z=\int P\big(\mathbf{y}(t)|\mathbf{x}(0),\boldsymbol{\Omega}\big)
P(\boldsymbol{\Omega})\,d\boldsymbol{\Omega},
\end{equation}
and $d\boldsymbol{\Omega}=d^N\mathbf{r}d^{N^2}\boldsymbol{\beta}d^N \boldsymbol{\sigma}$.
The first probability factor on the right-hand side of \eqref{eq:posterior} represents a product of log-normal distributions with width parameters $\boldsymbol{\sigma}$, centred at the trajectories $\mathbf{x}(t)$. Expressed in statistical language, the generative model behind the data is 
$$
y_k(t)\sim \text{log-Normal}\big(x_k(t),\sigma_k\big),
$$
where the abundances $x_k(t)$ are obtained by a simple numerical integration (e.g.~with a 4th-order Runge-Kutta method) of Eq.~\eqref{eq:lotka}. The last three probability factors in \eqref{eq:posterior} are the prior distributions of the model parameters.

Using the inferred posterior distribution of the parameters, we can sample new \emph{synthetic} trajectories $\hat y_k$, using the so-called \emph{posterior predictive} distribution
\begin{equation*}
P\big(\hat{\mathbf{y}}(t)|\mathbf{y}(t)\big)=\int
P\big(\hat{\mathbf{y}}(t)|\mathbf{x}(0),\boldsymbol{\Omega}\big)
P\big(\boldsymbol{\Omega}|\mathbf{y}(t),\mathbf{x}(0)\big)
\,d\boldsymbol{\Omega}.
\end{equation*}
Figure~\ref{fig:ppc} shows samples from the posterior predictive distribution for 6 noise levels. Note that the posterior predictive trajectories capture higher noise levels by increasing the variability for larger abundances.

To assess the results, in Sec.~S3 we collect the marginal posterior distributions for the model parameters (see Supplementary Information, Figs.~S1-S6 for 3 species, Figs.~S7-S12 for 4 species, and Figs.~S13-S18 for 4 species). For the rates $r_i$, we compute the quantile of the actual value used in the simulations. We consider good agreement between the inferred and the actual values if, for instance, that quantile is in the range 5--95\%. For the interaction parameters $\beta_{ij}$, we also show the quantiles and the probability that the predicted sign of each interaction coefficient is mistaken. This means that while the posterior-predictive trajectory accurately captures the data, it is compatible with a set of parameters that wrongly infer the sign.

To summarise these results, we define three key observables. First, the number of \textit{outliers}, which are parameters lying outside the 5th--95th percentile range. Second, the expected number of wrong signs (denoted by $\mp$ for simplicity), obtained as
\[
\mathds{E}\left[\mp\right] = \sum_{k=1}^{N^2} P_k(\mp),
\]
where $P_k(\mp)$ represents the probability that the $k$th element of the interaction matrix $(\beta_{ij})$ is estimated with the wrong sign. This probability is obtained by integrating the posterior for that parameter over the incorrect sign interval, namely, the area of the posterior probability for that parameter corresponding to a prediction having an incorrect sign. Finally, we estimate the overall probability that the model captures at least one wrong sign as
\begin{equation}
P(\mp) = 1 - \prod_{k=1}^{N^2} \left[1 - P_k(\mp)\right].
\label{eq:wrongsign}
\end{equation}

In Table.~\ref{tab:summary}, we show both metrics for the parameters corresponding to Fig.~\ref{fig:curve}(C) and additional examples for 3 and 5 species. Note how increasing the dimensionality produces more outliers for lower noise levels and increases the expected number of incorrect signs. This is not a failure of the method because Bayesian inference is the optimal way to quantify uncertainty in model predictions~\cite{jaynes2003probability}. The errors, quantified by the number of outliers and Eq.~\eqref{eq:wrongsign}, are not evident by inspection of the posterior predictive trajectories in Figs.~\ref{fig:ppc}, S19, and S20. Another conclusion is that there is an ensemble of parameters that explain the data but provide opposite interpretations, as the signs of $\beta_{ij}$ and $\beta_{ji}$ for every pair $(i,j)$ represent the type of interaction between species $i$ and $j$.

Finally, it is worth mentioning that the posterior distributions display strong correlations among parameters, although, as explained above, all growth rates, $r_i$, and interaction coefficients, $\beta_{ij}$, are randomly sampled. For instance, in Figs.~S22 and S23, we show a pairs-plot of the marginal posterior distributions. Note how correlations occur among different rows and columns of the interaction matrix. This suggests that aside from the concerns raised above---and summarized in Table.~\ref{tab:summary}---there is a strong redundancy in the model.

\begin{table*}[!htp]
\caption{Metrics quantifying how the posterior probabilities for the model parameters contain plausible sets of parameters that provide wrong predictions: $\mathds{E}[\mp]$, the expected value of the number of wrong signs; $P(\mp)$, the probability of predicting interactions with at least one wrong sign (relative to the true underlying value in the deterministic simulation); and the number of \textit{outliers}---parameters whose true value is outside the 5--95\% interval of the marginal posterior probability.}
\vspace{1mm}
    \centering
    \begin{tabular}{c|ccc|ccc|ccc} \hline
 \multirow{2}{*}{\textbf{Noise} } & \multicolumn{3}{c|}{\textbf{3 species}} & 
 \multicolumn{3}{c|}{\textbf{4 species}} & \multicolumn{3}{c}{\textbf{5 species}} \\ 
&$\mathds{E}\left[\mp\right]$
 &$P\left(\mp\right)$ & outliers& $\mathds{E}\left[\mp\right]$&$P\left(\mp\right)$& outliers& $\mathds{E}\left[\mp\right]$&$P\left(\mp\right)$& outliers\\\hline
0.001 & 0.00 & 0.000 & 1 & 0.05 & 0.050 & 3&  5.52&1 &     14\\
0.002 & 0.00 & 0.000 & 1 & 0.37 & 0.341 & 3&  6.25&1 &     14\\
0.003 & 0.00 & 0.000 & 0 & 0.73 & 0.624 & 4&  6.72&1 &     14\\
0.007 & 0.18 & 0.180 & 2 & 1.33 & 0.882 & 5&  7.58&1 &     13\\
0.010 & 0.50 & 0.500 & 3 & 1.82 & 0.938 & 5&  7.83&1 &     13\\
0.020 & 1.04 & 0.938 & 7 & 2.54 & 0.981 & 6&  8.47&1 &     12\\
0.030 & 1.47 & 0.984 & 7 & 2.76 & 0.986& 7 &  9.00&1 &     14\\
0.050 & 1.93 & 0.984 & 9 & 3.01 & 0.990& 8 &  9.76&1 &     14\\
0.070 & 2.56 & 0.996 & 9 & 3.39 & 0.992 & 9& 10.26&1 &     14\\
\hline
\end{tabular} 
\label{tab:summary}
\end{table*}

\begin{figure*}[!htp]
    \centering
     
    \begin{tabular}{ll}
        (A) 3 species &(B) 4 species \\
        \includegraphics[width=0.45\linewidth]{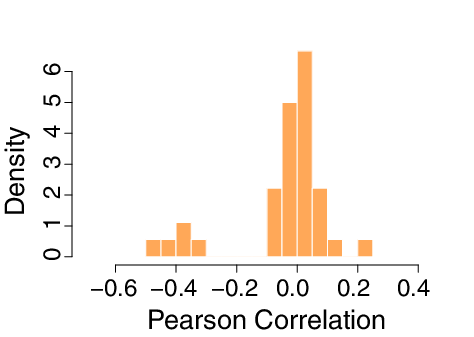}  
         &
         \includegraphics[width=0.45\linewidth]{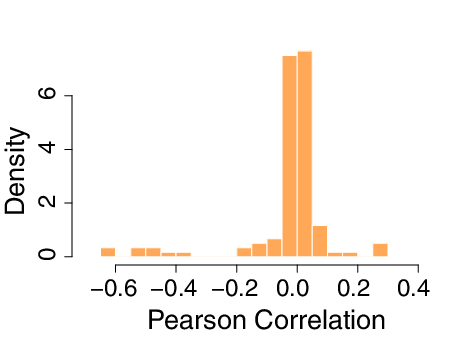}  
    \end{tabular}
    \caption{\label{fig:hist_corr} Normalised distributions of posterior marginal correlations between parameters for (A) the 3 species case (corresponding to Fig.~S22) and (B) the 4 species case (Fig.~S23). Note that, despite the original parameters being randomly sampled, inference mimics existing significant correlations (although most correlations are around 0). This is because there are sets of parameters that can be combined to reproduce the same reconstructed trajectory.}
\end{figure*}

To illustrate this redundancy, Fig.~\ref{fig:hist_corr} exhibits a normalised histogram of the pairwise correlations between the posterior marginals of the parameters, which correspond to Figures S22 and S23. Although the original parameters were sampled randomly, the inference reveals notable correlations that mimic existing significant relationships, even though most correlations cluster around zero. This occurs because the stability and feasibility conditions limit the distributions of the parameters compatible with the data, leading to effective correlations. In statistical language, the observed ecosystems represent a subsample of all the potential models \textit{conditioned} to produce feasible stable abundances.

\subsection{The gLV model is intrinsically sloppy}
\label{ssec:sloppy}

In the previous two sections, we have argued that typical population dynamics curves do not seem to contain enough information to estimate all the $N(N+1)$ parameters of the $N$-species gLV model. So far, we have shown that the posteriors entail the optimal use of information available in the data. Still, we have not provided a quantitative measure of what we mean by \textit{enough information} or how this can be exploited systematically. In this section, we apply the Manifold Boundary Approximation Method (MBAM) introduced in \cite{transtrum2014model}. This method uses information geometry \cite{Amari:2021} to characterise the $N(N+1)$-dimensional statistical manifold defined by the model parameters. The main idea is to introduce Fisher's information matrix  (FIM) as a metric tensor on this manifold and exploit its geometry to reduce the number of parameters with minimal information loss.

Generically, if our system is described by a log-likelihood $L(\mathbf{x}|\boldsymbol{\Omega})=\log P(\mathbf{x}|\boldsymbol{\Omega})$, where $\boldsymbol{\Omega}=(\omega_i)$ represents the parameters of the model, and if vector $\mathbf{X}$ represents the data to be fitted by the model, the elements of Fisher's Information matrix are given by
\begin{equation}
    I_{ij}(\boldsymbol{\Omega})=\mathbb{E}\left[\frac{\partial}{\partial\omega_i}L(\mathbf{X}|\boldsymbol{\Omega})\frac{\partial L}{\partial\omega_j}L(\mathbf{X}|\boldsymbol{\Omega})\right].
\end{equation}
The eigenvalues of this matrix quantify the change in the amount of information produced by a variation of the model parameters along the direction of the corresponding eigenvector. Thus, small eigenvalues characterise sloppy directions along which the model is still accurate enough, given the empirical data, but it can be driven to simplified sub-models containing as much information and preserving the ability to make accurate predictions with fewer effective parameters~\cite{brown2003statistical}. The idea of MBAM is to move along the geodesics defined by the metric tensor $I_{ij}$ toward the boundary of the manifold~\cite{gutenkunst2007universally}. The rationale for this is that geodesics are the straight lines of a curved geometry; hence, they represent the shortest paths to the boundary. For further details, see \cite{transtrum2014model}.

\begin{figure*}[!htp]
    \includegraphics[width=\textwidth]{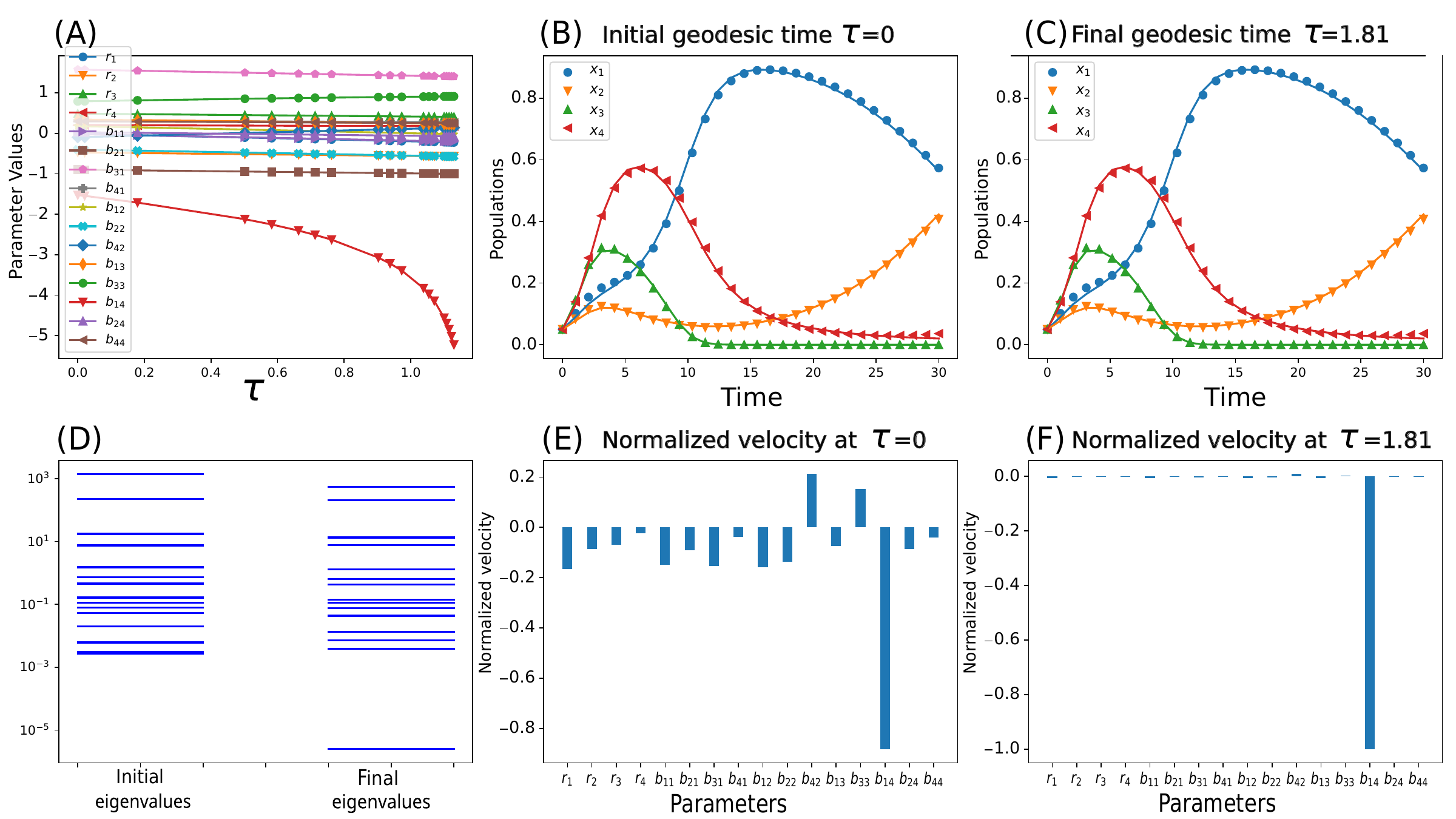}
    \caption{\label{fig:eigen_SSE}
    A combined plot to understand the MBAM. (A) trajectories of the parameters along the geodesics ($\tau$ being the intrinsic geodesic time, i.e., the parameter in which the geodesic curve is described). The other two panels in the first row show the data (symbols) and the model (solid lines) for the values of the parameters at (B) the beginning and (C) the end of the numerical integration interval. (D) Spectrum of Fisher's information matrix (FIM) for the parameters at the beginning (left) and end (right) of the integration interval. The smallest eigenvalue corresponds to the sloppiest combination of parameters. Notice how the sloppiest eigenvalue decreases upon reaching the manifold boundary through the geodesic. As parameters get removed, the spectrum shrinks within a few orders of magnitude. This is the signature of \emph{stiff} (non-sloppy) models. The criteria for determining when a manifold boundary is found are based on the rate of change (velocity) of the parameters along the geodesics. Panels (E) and (F) represent the velocity of the parameters at the beginning (E) and end (F) of the geodesic. The fact that just one parameter has a negative velocity at the end while the rest practically do not change means that the logarithm of this parameter tends to $-\infty$, and thus it can be dropped.}
\end{figure*}

We apply the MBAM systematically to another set of parameters of the gLV model, using a deterministic simulation with no added noise. This represents the (highly) ideal situation in which we possess perfect model knowledge. In Fig.~\ref{fig:eigen_SSE}, we illustrate one (arbitrary) step of the MBAM. In that figure, we show (A) the values of the logarithm of each parameter along the geodesic as a function of the geodesic time $\tau$ (geodesics are parametric curves with parameter $\tau$); the model fitted (B) at the beginning ($\tau=0$) and (C) at the end ($\tau>0$) of the integration interval for the geodesic; (D) the eigenvalues of the Fisher information matrix (FIM) at the beginning and end of that integration; and the velocity (rate of change with $\tau$) of each parameter along the geodesic (E) at the beginning and (F) at the end of the integration interval. In the particular case illustrated by the figure, $\log b_{14}\to -\infty$ as $\tau\to\infty$, meaning that this parameter is effectively 0.

\begin{figure*}[!htp]
    \centering
    \begin{tabular}{ll}
         (A)&(B)  \\
        \includegraphics[width=0.5\textwidth]{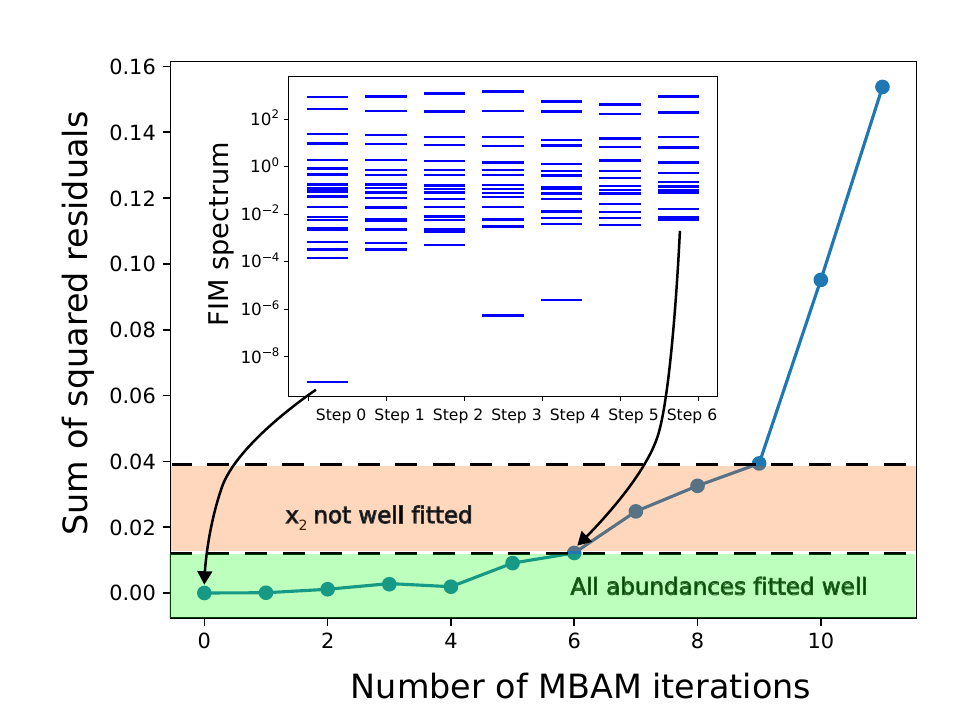}         
& 
         \includegraphics[width=0.5\textwidth]{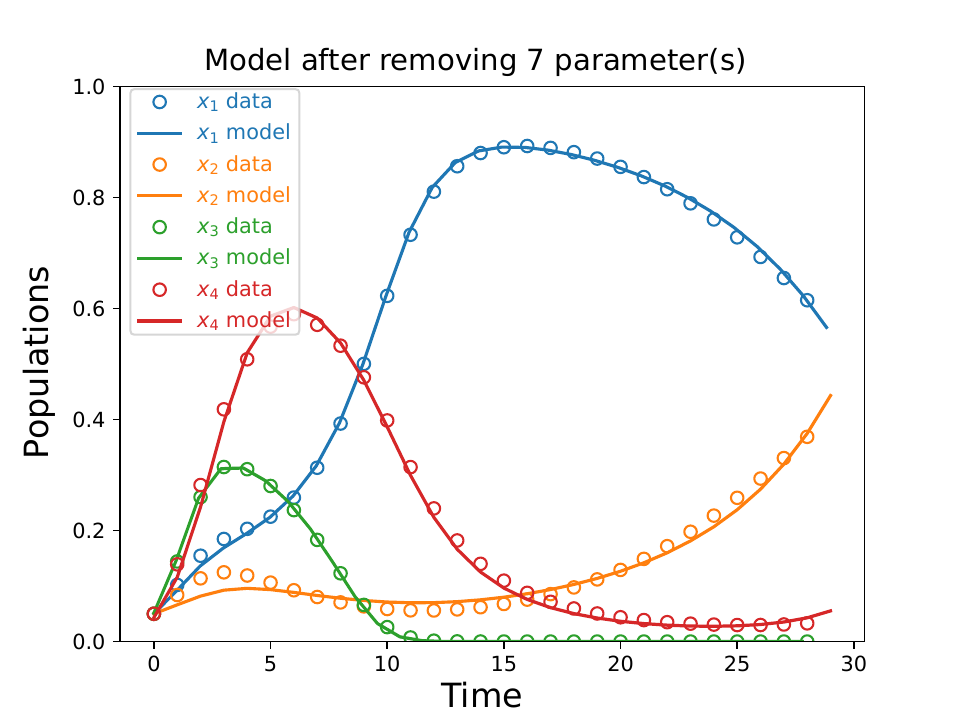}   
    \end{tabular}
    \caption{\label{fig:sse} (A) Sum of squared residuals (difference between original data and model prediction) for model parameters in Fig.~\ref{fig:eigen_SSE}. Inbox: MBAM parameter elimination shrinks the FIM spectrum, making the resulting model less sloppy according to Ref.~\cite{transtrum2015perspective}. 
    Note how the error jumps at 7 reduced parameters corresponding to the inability of the reduced model to explain $x_2$, as shown in (B).}
\end{figure*}

We computed 11 steps of MBAM. In the first 6 of them, we can eliminate 6 parameters safely ($\beta_{32}$, $\beta_{23}$, $\beta_{34}$, $\beta_{43}$, $\beta_{14}$, and $\beta_{42}$). In the next 3 steps, we can eliminate 3 more parameters ($\beta_{22}$, $\beta_{33}$, and $\beta_{21}$) if we are willing to sacrifice the goodness of fit of species $x_2$. Finally, the last 2 steps illustrate how the fitting errors rise dramatically in a much too simplified model. Figure~\ref{fig:sse}(A) shows the evolution of a measure of the fitting error (the sum of squared residuals) after each MBAM step. Note in Fig.~\ref{fig:sse}(B) how, after removing the 7th parameter, the trajectory of $x_2$ is no longer well captured. The inset in Fig.~\ref{fig:sse}(A) shows how the FIM spectrum shrinks by reducing the complexity of the model, corresponding to a less sloppy model after each step. Note also that the spectrum of the FIM for the final reduced model (after 6 steps) now spans just a few orders of magnitude, meaning that the model is no longer sloppy. 

Another suggestive conclusion from this systematic reduction is that MBAM has eliminated all the parameters by making them 0. Hence, the interaction matrix of the reduced model is, after reduction, sparser. This sparsity has been recently pointed out as a signature of microbial ecosystems~\cite{camacho2024sparse}. Note that removing parameters by making them 0 is not the only outcome of the MBAM (which also accommodates simultaneously pushing parameters to infinity). Still, we speculate that it might be a consequence of the mathematical structure of the gLV model.

In Fig.~S24, we show an example resembling the typical cycles in the original predator-prey formulation of the Lotka-Volterra model. As in the previous example, all the MBAM steps eliminate interaction coefficients.

Note that when using MBAM (see e.g.~\cite{transtrum2014model}), the simplifications we achieve are contingent on the data we are fitting. Having more \emph{relevant} information makes further model parameters relevant, and the oversimplified model here obtained may be too simple. By relevant information, we do not mean necessarily longer times (as shown in Sec.~S8) but truly new information not present in the previous data (typically, new experiments performed with different conditions or measuring additional observables).

By the same principle, it is worth emphasising that all these numerical experiments are made with noiseless, synthetically generated data so that they would correspond to experimental data measured with extreme accuracy. Combined with the discussion about practical identifiability emerging from the Bayesian analysis in Sec.~\ref{ssec:bayes}, we expect the reduction to be even more dramatic in a real scenario. Moreover, taking into account that we have more sampling points in our time series than typically available in experiments, this expectation is reinforced.

\section{Discussion and conclusions}
\label{sec:discusion}
Our overarching argument throughout this paper is simple but has deep implications: population dynamics time series do not contain enough information to estimate and confidently interpret the interactions between species in an ecosystem. Typically, the time evolution is governed by a few timescales and typical abundance sizes (what we summarise in Fig.~\ref{fig:curve}(A) as \textit{pieces of information}). This hand-waving assessment---which we made more quantitative through Eq.~\eqref{eq:generic}---has two implications: one regarding the extent to which we can infer interactions from the data (using Bayesian inference, the optimal way to extract information from data), and another one using the principles of information geometry (which provides a systematic way to remove non-informative parameters).

Regarding the first, we showed (through the marginal posterior distributions for the parameters) that, although the problem is in principle structurally identifiable~\cite{remien2021a}, in practice, there is a significant probability that not only the value but even the sign of a given interaction parameter $\beta_{ij}$ is wrong---unless the priors are so close to the original parameters that the posteriors are biased beforehand (see Sec.~S7). This has important biological implications because we can incorrectly predict the ecological nature of species interactions (competition, commensalism, mutualism, parasitism\ldots). Also, the posterior densities show that the real parameter used in the simulation cannot be practically identified due to noise. Furthermore, the larger the population is, the more uncertain determining the underlying interactions (even their sign) becomes.

On the other hand, MBAM illustrates how the gLV model is over-parametrised, thus substantiating our graphical argument in Fig.~\ref{fig:curve}(A). The first conclusion is that the removed parameters are, mainly, coefficients of the matrix $\beta_{ij}$ set to 0. This aligns with the idea, recently put forward in a study that infers interaction matrices of microbial communities~\cite{camacho2024sparse}, that the interaction matrices  $\beta_{ij}$ of these systems tend to be sparse. Of course, all our analyses have been performed on relatively smooth time series. Whether the argument still carries on to more general cases (with many more species) involving complex oscillations or chaotic behaviour needs further study.

Another implication of our results is that, as gLV (or more sophisticated models) are overparametrised, they are expected to be more prone to overfitting and to generalise poorly to future events, with this effect being larger for a larger number of species. We illustrate this in Sec.~S8. Interestingly, extending the model reduction up to $t=6$ improves the fitting, but the inferred (and eliminated) parameters differ, as shown in Fig.~S27.

One should bear in mind that the problems revealed in inferring the interactions of the gLV model are not exclusive to this system but apply to any sufficiently complex dynamics \cite{gutenkunst2007universally, transtrum2015perspective}. Thus, our discussion here is completely general and pertains to the science of complex systems at large.

Any critical reader might suggest potential solutions to this pessimistic scenario. In the problem under discussion, one possible solution would be to extend the gLV model to include additional mechanisms---such as competition/production for metabolites or other resources~\cite{skwara2023statistically}. This might indeed improve our understanding of microbial interactions, but those models have a complexity cost that should be supplemented with relevant empirical data. Otherwise, our main argument about the relevant \textit{pieces of information} in the data will imply that the new parameters related to the species-resource interplay will suffer from the same problems as those described in this paper.

Alternatively, we can employ stochastic versions of the gLV model \cite{descheemaeker2020stochastic, camacho2024sparse}. These generalisations offer additional insights by accommodating variability in the data as random fluctuations. This helps reduce overfitting and suggests focusing on the distributions of interactions. However, implementing this method is not straightforward, as the stability and observability of an ecosystem already influence the underlying correlations within the data (as illustrated in Fig.~\ref{fig:hist_corr}).

Instead of approaching the problem by trying to amend the model, we advocate for a paradigm shift in ecological research, moving the focus from traditional dynamical systems to a framework grounded in the principles of Statistical Mechanics. The central challenge lies in defining the appropriate state variables that capture the essence of ecological dynamics and facilitate specific strategies for ecosystem manipulation toward desired outcomes. We introduce an ensemble approach to the analysis of ecological data, positing that the joint posterior distribution, $P\big(\boldsymbol{\Omega}|\mathbf{y}(t),\mathbf{x}(0)\big)$, extends beyond mere statistical inference. Specifically, instead of referring to a parameter as the unique numerical value that better fits the data, this approach advocates for assigning a whole ensemble of parameters compatible with the data, so the interpretation of the model parameters is more robust; on the one hand, and contains additional information about the uncertainty associated to that parameter, on the other hand. Thus, this ensemble approach serves as a comprehensive description of the ecological system in a manner akin to the Maxwell distribution that characterises the velocities of gas molecules.

In conclusion, this framework encompasses a multitude of potential scenarios that can explain observed data, thereby maximising the informational yield from experimental findings. Accordingly, it is essential to develop robust macroscopic state variables that encapsulate critical ecological phenomena such as diversity, stability, symbiosis, and interspecies competition, all of which can be derived from these joint distributions. As we look to the future, we recognise that this presents an exciting and necessary research agenda that promises to deepen our understanding of ecological systems and enhance our ability to manage and sustain them effectively.

\end{multicols}

\enlargethispage{20pt}

\ethics{No ethical declaration required.}

\dataccess{Data is generated automatically from the code available at   \href{https://zenodo.org/records/16747311}{https://zenodo.org/records/16747311}}

\aucontribute{ All the authors contributed equally except MC, who performed the numerical simulations, Bayesian inference, and MBAM model reduction.}

\competing{No competing interests.}

\funding{This work has been supported by grants PID2021-122711NB-C21, PID2022-141802NB-I00 (BASIC), PID2021-128966NB-I00, and PID2022-140217NB-I00, funded by MICIN/AEI/10.13039/501100011033 and by ``ERDF/EU A way of making Europe''.}

\ack{We thank Susanna Manrubia and Sa\'ul Ares for fruitful discussions about minimal models and parameter interpretation in sloppy systems, and Aniello Lampo for critically reading the manuscript and useful suggestions. }

\disclaimer{No disclaimers.}

\printbibliography

\end{document}